\documentclass{article}
\usepackage[a4paper, margin=1in]{geometry}
\usepackage{amsmath}
\usepackage{xurl}
\usepackage{graphicx} 
\usepackage[dvipsnames]{xcolor}
\graphicspath{ {./figures/} }
\usepackage{setspace}
\usepackage{array}
\title{Perception of dynamic multi-speaker auditory scenes under different modes of attention}
\author{Stephanie Graceffo, David F Little, Emine Merve Kaya, Mounya Elhilali\thanks{Corresponding author: mounya@jhu.edu} \\ Laboratory for Computational Audio Perception\\ Department of Electrical and Computer Engineering \\ Johns Hopkins University, Baltimore, MD, USA}
\date{}

\begin{document}

\maketitle

\section{Abstract}
Attention is not monolithic; rather, it operates in multiple forms to facilitate efficient cognitive processing. In the auditory domain, attention enables the prioritization of relevant sounds in an auditory scene and can be either attracted by elements in the scene in a bottom-up fashion or directed towards features, objects, or the entire scene in a top-down fashion. How these modes of attention interact and whether their neural underpinnings are distinct remains unclear. In this work, we investigate the perceptual and neural correlates of different attentional modes in a controlled "cocktail party" paradigm, where listeners listen to the \emph{same} stimuli and attend to either a spatial location (feature-based), a speaker (object-based), or the entire scene (global or free-listening) while detecting deviations in pitch of a voice in the scene. Our findings indicate that object-based attention is more perceptually effective than feature-based or global attention. Furthermore, object-based and spatial-based attention engage distinct neural mechanisms and are differentially modulated by bottom-up salience. Notably, while bottom-up salience aids in the initial segregation of auditory objects, it plays a reduced role in object tracking once attention has been voluntarily allocated. In addition, decoding the stimulus envelope from the EEG data revealed a source-sampling scheme in the global attention mode that is not present in the object or spatial modes. Overall, the study shows that the perception of the same acoustic scene differs according to the listening task, guided by an interaction between top-down and bottom-up processes.

\section{Introduction}

To cope with limited cognitive resources available, the brain deploys attention as adaptive mechanisms to sort and prioritize sensory information. These processes direct focus towards relevant stimuli while filtering out distractions. In the auditory domain, attention enables us to navigate complex sound environments by making sense of competing sources that impinge on our ears. The cocktail party effect \cite{Cherry1957, Haykin2005} illustrates this ability, allowing us to follow a conversation of interest despite background noise and select sounds of interest. Long standing debate centers on \emph{where} in the sensory hierarchy that selection is enacted and \emph{what} unit of representation it targets. One influential hypothesis argues for an object centered view: Attention locks onto a perceptual object (e.g., a single voice) and strengthens \emph{all} its features as a coherent package, allowing effortless tracking across pitch fluctuations or movement of the sound across space \cite{Shinn-Cunningham2008,Kerlin10a,Mesgarani2012}. A competing account emphasizes feature selective mechanisms, proposing that attention can be steered toward low level attributes such as spatial direction or fundamental frequency, with ‘objecthood’ emerging only after an attended feature has been isolated \cite{Zatorre1999, Petkov2004}. Reconciling these views requires clarifying whether feature based and object based selection are separable processes or simply different entry points into a shared hierarchy.

Evidence from vision underscores the complexity of this issue. Spatial, feature, and object attention each leave distinct neural fingerprints \cite{Treue1999,Saenz2002,Liu2011a}, yet they interact in non trivial ways. Object-centric theories predict a spread of attention, whereby focusing on one feature of an object automatically boosts its unattended attributes \cite{OCraven1999, Schoenfeld2003}. By contrast, more recent work shows that feature based selection can \emph{suppress} other object features when they are task irrelevant \cite{Wegener2008, Freeman2014}. To reconcile these seemingly contradictory results, current theories of visual processing suggest that feature- and object-based attentional mechanisms coexist in the visual pathway and operate as parallel filters that can cooperate or compete depending on task demands \cite{kravitz2011,Mayer2012,Wegener2014,Geigerman2016}. 

Auditory research mirrors this debate but with less consensus. Early PET and EEG studies reported little evidence for feature specific gain, concluding that attention targets integrated auditory objects (e.g. an entire voice) rather than low level cues like frequency or spatial hemifield \cite{Zatorre1999,Petkov2004}. Other studies including fMRI, MEG and intracranial recordings demonstrated robust enhancement of tonotopic and spatial maps when listeners attend to a single attribute, supporting a genuinely feature based mechanism in both primary and non primary cortex \cite{Ahveninen2006,Krumbholz2007,dacosta2013}. Crucially, suppression of an unattended feature which is diagnostic of independent feature filters has yet to be unambiguously demonstrated in audition \cite{Paltoglou2009}. Thus, two interpretations remain viable: (i) feature attention is simply an early manifestation of object centered processing, or (ii) feature and object selection are distinct but cooperative processes that modulate auditory cortex in parallel. A critical test, therefore, is to present identical acoustic scenes while instructing listeners to adopt object, feature,  or scene level goals, and to compare both perceptual efficacy and neural signatures across these modes.

Attention without explicit goals, sometimes labeled global, free listening or non selective attention, is yet another form of attention that has received far less scrutiny. Visual work suggests that global attention yields rapid ``gist’’ extraction of scene layout before finer analysis \cite{Navon1977,Oliva2007} and recruits early, low spatial frequency channels \cite{hedge2008}. In audition, mismatch negativity studies show that deviant sounds can be detected even when they are task irrelevant, implying a parallel global monitor \cite{Naatanen2007}. Whether that monitoring filter integrates the whole scene into a unitary representation or, alternatively, \emph{samples} individual sources in a serial fashion is still debated. One hypothesis is that, absent task goals, attention cycles through salient sources, guided by bottom up conspicuity and exploratory drive \cite{Kaya2014}. A counter hypothesis posits a spatial scaffold: even in global mode, attention defaults to stable spatial anchoring, enhancing events that follow the prevailing layout \cite{bohm2013}. Distinguishing these models is vital for understanding everyday listening, where listeners frequently drift between focused and diffuse states.

Another unresolved dimension concerns top down/bottom up interplay. When a listener locks onto a target, does increasing the intrinsic salience of a competing stream break that lock (a capture model), or are salient distractors actively suppressed once selection is established (a gating model)? Conversely, in global listening, is salience beneficial because it guides the sampling schedule or is it detrimental because simultaneous salience bursts create mutual masking? Prior studies rarely manipulate salience and attentional modes orthogonally, leaving the field with contradictory predictions.

\begin{figure*}
    \centering
    \includegraphics[width=\textwidth]{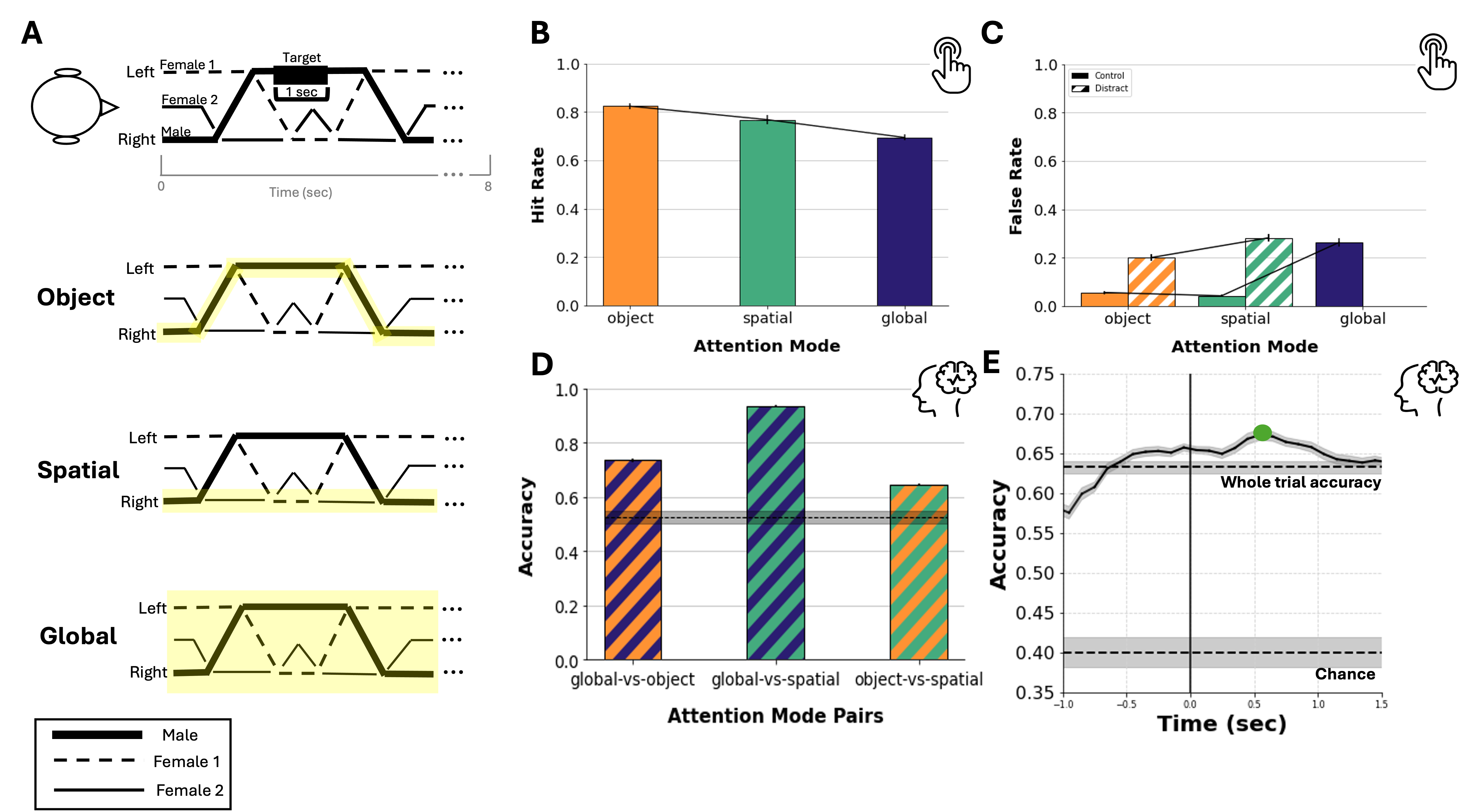}
    \caption{\textbf{(A)} Setup of the cocktail-party stimulus used in this experiment. The stimuli consists of book sections narrated by three German speakers.  The three speakers start at the left, center, and right, with occasional shifts in each direction. Targets are 1 second long segments of natural speech manipulated to have a pitch outside the pitch range of regular speech (3 semitones lower for male targets and 3 semitones higher for female targets). \emph{\textbf{Lower}} Subjects listened to each auditory scene three times: Object (top) by attending to the male speaker regardless of spatial location, spatial (middle) by attending to the voices on the right, and global (bottom) by attending to the entire scene. \textbf{(B)} Target detection performance during each of the three attention modes of the behavioral experiment.  Global attention represents free-listening, providing a baseline to evaluate selective attention.  Directing attention to an acoustic spatial location (right side) and object (male speech) results in progressively increasing hit rates. Bars represent average hit rates for each condition and error bars represent standard error of the mean. Statistical significance is reported as (*), $p < 0.05$, (**) $p < 0.01$, (***) $p < 0.001$, (NS) Not Significant. The icon on the upper-right corner highlights that results reported in figure are based on behavioral responses. \textbf{(C)} False positive responses are elicited from both control trials (trials with no targets) and distraction trials (trials with a target in the unattended stream).  \textbf{(D)} Neural responses from subjects elicit different signatures under each attention mode that are distinct enough for successful classification. Average accuracy for pair-wise classifiers is reported across different cross-validation runs. The gray line represents the accuracy of chance classifier obtained from randomized labels. The icon on the upper-right corner highlights that results reported in figure are based on neural responses. \textbf{(E)} three-way classification of a 0.75-second window beginning 0.25 seconds before the target onset shows a slight boost in accuracy compared to classifying across the whole trial}
    \label{fig:attention_modes}
\end{figure*}

To sift through these competing views, the current study employs identical, spatially dynamic “cocktail party’’ scenes and ask them to adopt three attentional modes in separate blocks (Figure \ref{fig:attention_modes}A): (1) monitor the entire mixture (\emph{global}), (2) attend to any speech on the right (\emph{feature/spatial}), or (3) track the male voice wherever it appears (\emph{object}). Three voices (a male and two females) constitute the scene and all three move dynamically in space but are never concurrently present in one spatial location at the same time. All narrations are in German, a language unfamiliar to our listeners hence engaging the auditory system at a more sensory level and relying less on linguistic processing. Targets (pitch shift) at carefully positioned moments in the scene facilitate monitoring of attentional focus of participants. This design allows us to examine explicit hypotheses:
\begin{itemize}
\item[H1] - Dominance: Object-based attention will outperform feature-based attention, suggesting an object first hierarchy. A counter hypothesis is that feature-based attention will be equally effective once the feature is diagnostic and sufficient for segregation and tracking, implying shared mechanisms.

\item[H2] - Global sampling: Global attention operates via serial source sampling and neural markers will cycle across voices. A counter hypothesis is that global attention preserves a spatial map and that neural activity will remain anchored to spatial layout of the entire scene rather than individual streams.

\item[H3] - Salience modulation: Bottom up salience boosts performance primarily in global mode but contributes little once selective attention is engaged. A counter hypothesis is that highly salient events capture attention even within selective streams, resetting the attentional focus.
\end{itemize}

We evaluate these competing accounts using converging behavioral metrics, stimulus reconstruction decoding and spectro-temporal EEG patterns. Our approach provides a framework to delineate when and how auditory attention privileges objects, features, or exploratory sampling in a dynamic setting mimicking real world listening.

\section{Results}

\subsection{Different modes of attention yield distinct behavioral performances}
Even though the stimuli are exactly the same, behavioral responses reveal that the three modes of attention are not equally effective.  Figure \ref{fig:attention_modes}B shows a clear ordinal pattern in hit rates, with global attention resulting in the lowest performance (HR = 0.695 \textpm 0.012), followed by spatial attention (HR = 0.768 \textpm 0.019), and object attention with the highest performance (HR = 0.824 \textpm 0.013).  A Friedman test comparing the effect of hit rates on the three attention modes reveals a statistically significant difference in hit rate and attention modes (F=55.2, p=1.01e-12).  A Wilcoxon signed-rank test with Bonferroni corrections between hit rates shows a significant difference between global and object (W=100, p=9.57e-12), global and spatial (W=395, p=7.05e-7), and object and spatial (W=520, p=8.24e-4) modes. 

Incorrect responses in this experiment stem from two sources of false positives: control trials and distraction trials.  Control trials do not contain a target, while distraction trials contain a target in the unattended stream.  This latter case occurs when subjects are performing the object task and there is a target in the female voice, or subjects are performing the spatial task and there is a target on the left side. Figure \ref{fig:attention_modes}C shows that false positive responses to control trials (solid bars) are prominently smaller for spatial and object attention (FP = 0.042 \textpm 0.006 and FP = 0.056 \textpm 0.007, respectively) compared to global attention (FP = 0.263 \textpm 0.018).  Wilcoxon signed-rank tests with Bonferroni corrections between false positive rates reveal a significant difference for global and spatial (W=184, p=7.25e-6), and global and object (W=174.5, p=7.8e-4), but not for object and spatial modes (W = 189.5, p = 0.096). In object and spatial attention tasks, distraction trials contain targets in the unattended stream (striped bars in Figure \ref{fig:attention_modes}C).  The false positive rate for the spatial mode (FP = 0.282 \textpm 0.016) is significantly higher (W=85.5, p=1.72e-6) than that of the object mode (FP = 0.200 \textpm 0.014). Furthermore, comparing distraction to control rates for each mode confirms significantly higher distraction false rates (object mode (W=162.0, p=6.47e-12), spatial mode (W=19.0, p=1.64e-13)). Overall, the higher hit rate and lower distraction rate for the object task suggest that object-based attention is more perceptually effective than spatial attention for scenes that involve dynamic spatial configurations. Furthermore, the significantly higher distraction rates compared to control rates in both spatial and object attention allude to a failure of attentional deployment to the correct target rather than a failure in segregation. 

Behavioral sensitivity across the three attention modes -measured using $d'$- follows the general trend of hit rates. Global attention has the lowest value ($d'$ = 1.31 \textpm 0.826), followed by spatial ($d'$ = 1.42 \textpm 0.845), then object attention ($d'$ = 1.84 \textpm 0.850). A Friedman test comparing the effect of all three attention modes on $d'$ reveals that there is a statistically significant difference in $d'$ and attention modes (F=24.11, p=5.812e-6).  Wilcoxon signed-rank test with Bonferroni corrections between $d'$ values reveals a significant difference in $d'$ values for global and object (W=574, p=3.29e-5) and for object and spatial (W=549, p=1.77e-5) but not for global and spatial modes (W=1225, p=0.617). The higher $d'$ for object attention compared to spatial attention supports the reasoning that auditory object tracking across spatial locations is facilitated by object-based attention but hindered during spatial-based attention due to the dynamic spatial nature of the scene.

The same analyses are replicated for behavioral responses obtained concurrently with neural recordings (albeit it with a smaller sample size) compared to the behavioral only experiment. Comparison of the two responses using permutation tests (see Methods) shows no significant difference between hit rates (object p=0.589, spatial p=0.162, global 0.300), false positive rates (object p=0.958, spatial=0.791, global=0.038), or $d'$ (object p=0.855, spatial=0.958, global=0.061).

\subsection{Distinct neural signatures underlie different modes of attention}
\label{section:classifier}
Neural responses to dynamic scenes under different modes of attention result in different EEG signatures, clearly distinguishable from one another (Figure \ref{fig:attention_modes}D).
A classifier is trained on EEG responses from trials where subjects correctly detect the target, reasoning that these trials reflect proper attentional engagement aligned with task demands for each block. Classification accuracies for pairwise comparisons are highest for global vs spatial (93.5 \textpm \ 0.03\%), then global vs object (73.6 \textpm \ 0.71\%) and finally object vs spatial (64.4 \textpm \ 0.46\%) modes.  All attention mode pairs perform well above chance (52.6 \textpm \ 2.32\%), derived using a shuffling procedure. Furthermore, a three-way classifier comparing the three modes of attention yields an average accuracy of 63.2 \textpm \ 0.72\% across the entire trial (chance accuracy 40.0 \textpm \ 1.87\%). One of the challenges in the experimental design is that each trial follows a distinct spatial layout, complicating the alignment of analyses across trials. Nonetheless, aligning the analysis near onset of target events reveals a slight, yet significant boost in classification accuracy about 500ms after target onset relative to whole-trial accuracy (68.8 \textpm \ 0.64\% relative to 63.2 \textpm \ 0.72\%, t-test $p=3.22e-5$).

\subsection{Scalp topographies support distinct neural activity patterns under different modes of attention}
\label{section:heatmaps}

\begin{figure*}
    \centering
    \includegraphics[width=\textwidth]
    {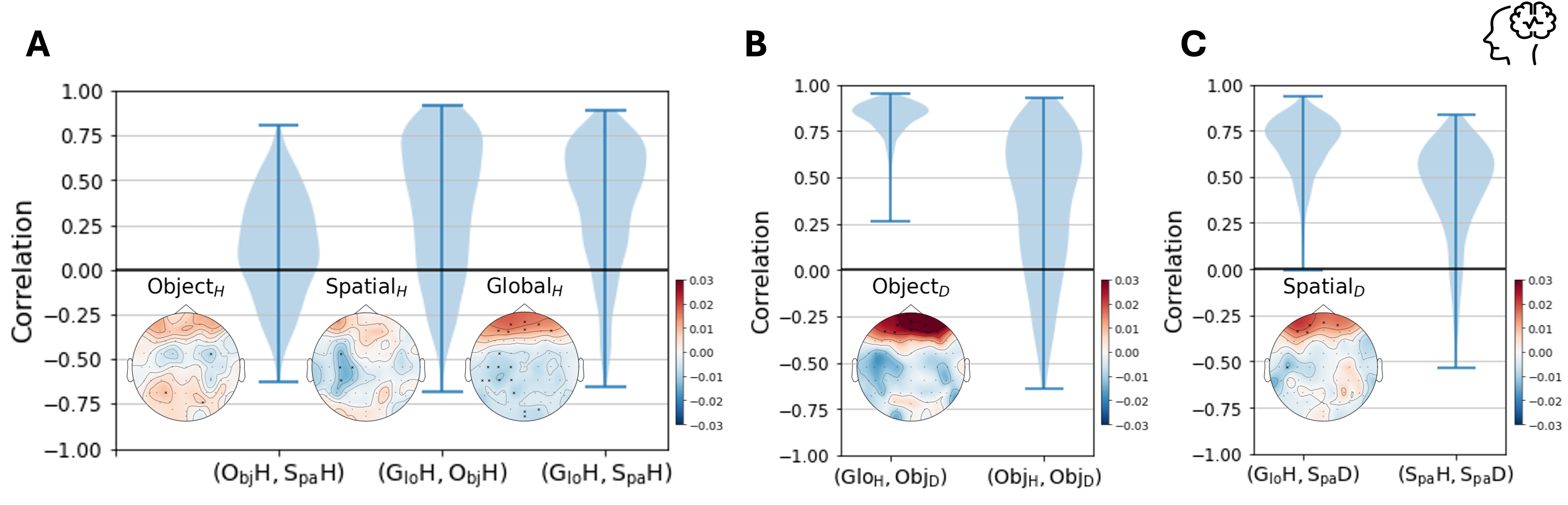}
    \caption{\textbf{(A)} \textit{Hit Responses}: Correlations between topographies for hit trials for attention mode pairs are not significantly different from zero indicating low similarity between pairs.  The inserted topographic maps of the object, spatial, and global attention modes qualitatively illustrate these differences as well.  \textbf{(B)} \textit{Object False Positive Responses}: Object false positive and global hit responses are significantly correlated, while object hit and object false positive responses are not significantly correlated.  This can also be observed qualitatively when comparing the scalp topographies of the object false positive to those of the global hit and object hit in A. \textbf{(C)} \textit{Spatial False Positive Responses}: Spatial false positive and global hit responses are significantly correlated, while spatial hit and spatial false positive responses are not significantly correlated.  This can also be observed qualitatively when comparing the scalp topographies of the spatial false positive to those of the global hit and spatial hit in A.}
    \label{fig:heatmaps}
\end{figure*}

Figure \ref{fig:heatmaps}A depicts heatmaps in the three attentional modes during hit trials allowing us to examine possible underpinnings of differences between attention tasks. Average scalp topographies reveal distinct patterns in the three tasks. The object attention condition shows the expected frontal and bilateral occipital activation though accompanied by a weak temporal activation; while spatial attention reveals a stereotypical parietal-occipital suppression contra-lateral to the attended location with moderate frontal activation. In contrast, the global mode yields a very strong frontal enhancement with general posterior parietal suppression, likely to enhance auditory processing. An analysis of neural oscillations that may drive these topographic differences is performed across different frequency bands and anatomically distinct channel clusters. A one-way ANOVA comparing theta activity in frontal channels across the three attention modes reveals a significant main effect (F(2, 46) = 7.35, p = 0.002). Post-hoc paired t-tests with Bonferroni correction yields a significantly higher theta power in global vs. spatial (t(23) = 3.12, p = 0.005) and object vs. spatial (t(23) = 2.85, p = 0.009) and even higher in global vs object condition (t(23) = 4.21, p = 0.001). In addition, spatial attention shows a significantly higher parietal alpha power relative to both object (t(23) = 2.02, p = 0.011) and global (t(23) = 2.02, p = 0.011) modes.

To further compare the overall patterns across all channels, a pair-wise correlation of different attention modes using bootstrapping is performed (see Methods). Figure \ref{fig:heatmaps}A reveals that comparing heatmaps derived during correct hit trials yields correlations that are not significantly different than zero, indicating a low similarity and supporting that neural activity is unique for each mode of attention. This analysis is then replicated to probe parallels between \emph{distraction trials} in selective attention modes (object and spatial) and correct trials in global attention. An interesting trend emerges when comparing the distraction trials for object and spatial modes to their respective hit responses in selective tasks and to hit responses in the global attention task.  Figure \ref{fig:heatmaps}B shows that the object false positive distract trials are significantly correlated with the global hit trials with 95\% CI [0.32, 0.90] but not with the object hit trials, 95\% CI [-0.30, 0.83].  Notably, the average scalp topography for the object false positive distract trials are qualitatively aligned with the global hit trials with stronger frontal activation and more posterior parietal suppression. Similarly, Figure \ref{fig:heatmaps}C shows that the spatial false positive distract trials are significantly correlated with the global hit trials with 95\%CI [0.20, 0.87], but not with the spatial hit trials, 95\% CI [-0.30, 0.77].  Furthermore, the average topography for the spatial false positive distract trials (Figure \ref{fig:heatmaps}C insert) also more closely resembles that of the global hit trials (Figure \ref{fig:heatmaps}A insert) than it does the spatial hit trials (Figure \ref{fig:heatmaps}A insert), particularly in terms of strong frontal activation. In the case of spatial distract trials, we still note a mild contra-lateral suppression.

\begin{figure*} 
    \centering
    \includegraphics[width=\textwidth]{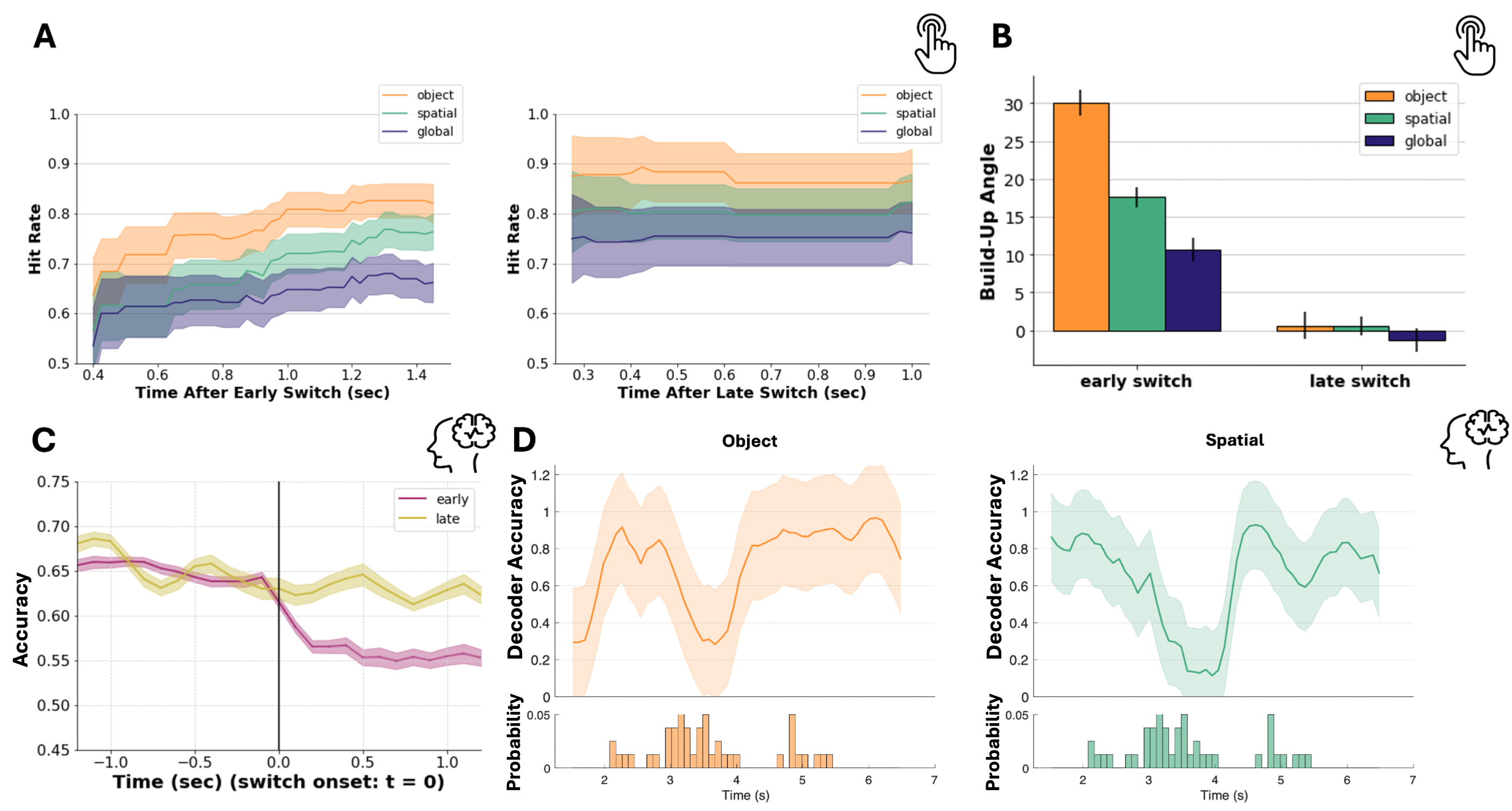}
    \caption{\textbf{(A)} Time course of hit rate for early (before or after the first speaker switch) attention reorientation (\textit{Left}) and for late (after the second or third speaker switch) attention reorientation (\textit{Right}). \textbf{(B)} Build-up angle of the time course for early and late shifts.  Angle is found by fitting a line at the build-up time window.  \textbf{(C)} 3-way classification as a function of time from switch onset for the early and late switch \textbf{(D)} Decoding accuracy for Object with probability distribution for times where speakers converge on the same side (\textit{Left}). Decoding accuracy for Spatial with probability distribution for times where speaker moves away from the attended (right) side (\textit{Right}).
    }
    \label{fig:temproal_dynamics}
\end{figure*}

\subsection{Different attention modes follow distinct temporal patterns influenced by scene dynamics}

To explore temporal dynamics of distinct attention modes, behavioral and neural responses as a function of time are examined relative to the spatial dynamics of the scene in each trial. Figure \ref{fig:temproal_dynamics}A shows the hit rate as a function of time after the switching point (the time at which the speakers change directions, moving from one location in space to another) and reveals different trends of build-up as the scene unfolds.  Rates at every time $t$ denote the average hit rate of targets that are occurring at that time.  For targets that occur ``early'' in the trial (after trial onset or following an early spatial shift), the hit rate build-up lasts over 100's of milliseconds (up to 1.2 seconds after the shift) before stabilizing. Surprisingly, targets that happen after a ``late'' shift do not show this pattern; instead, all three attention modes show a stable hit rate level over time. 

Quantifying this buildup trend by estimating the angle of the slope of the hit rate curve using a linear fit confirms that build-up angles are most prominent for the early shifts, but there is virtually no build-up observed for the late shifts (Figure \ref{fig:temproal_dynamics}B).  A two-way ANOVA to examine the effects of attention mode and switch time on hit rate confirms a significant main effect of attention mode on hit rate (F=148.9, p=1.0e-43), indicating that performance differed across attention modes.  There is also a significant main effect of switch time on hit rate (F=448.0, p=2.51e-58), with higher performance for the late switches.  However, there is no significant interaction between attention mode and switch time (F=0.62, p=0.538), suggesting that the effect of switch time on performance is consistent across tasks.

It is worth noting that although all modes have stabilized, the object hit rate is still larger than the spatial hit rate, which are both larger than the global hit rate, once again illustrating that object attention is stronger than spatial attention, and both types are stronger than global attention.

Examining the neural responses around the time of scene switches also reveals different trends in brain responses to the three attentional modes, reflecting different effects of attention reorienting. Figure \ref{fig:temproal_dynamics}C shows classification accuracy of neural responses under the three tasks as a function of time, aligned near the onset of the switch. The results show that a three-way classifier is generally stable before the onset of the switch (average classification accuracy for 1 second before is 62.8 $\pm$ 4.0\% for early switches and 65.1 $\pm$ 2.5\% for late switches). In contrast, the classification accuracy drops to 54.8 $\pm$ 1.2\% after early switches while it remains stable at 62.3 $\pm$ 2.7\% for late switches. This pattern aligns closely with that observed behavioral responses and suggests that attention reorienting occurs more prominently earlier in the trial as voices shift in the scene before stabilizing as the spatial configuration changes become part of the regularity in the scene.

\subsection{Distinct temporal dynamics characterize selective attention to complex auditory environments under object and spatial modes}
\label{section:decoding_target_other}

To further investigate how attention unfolds over time, we decode the stimulus envelope from the neural data, and examine the decoding correlation (between actual and reconstructed envelopes) as a function of time for each selective attention mode. This analysis focuses on the object and spatial tasks as they both have a clear notion of ``attended'' versus ``unattended'' perceptual streams. Separate decoders are trained for each voice and each spatial location in the stimulus and are grouped into attended stream decoders (male left and male right for the object task; male right, female 1 right, and female 2 right for the spatial task) and unattended stream decoders (see section \ref{section:decoder}). The decoding accuracy is denoted as 1 if the attended output exceeds the unattended output and 0 otherwise.

Figure \ref{fig:temproal_dynamics}D shows the average decoding accuracy over time for the object and spatial conditions.  For the Object condition (left), the probability histogram below the curve represents the probability of one speaker starting to switch towards the side of the auditory scene where another speaker already is (speaker convergence). This condition captures increased segregation difficulty given that multiple speakers are likely to share the same spatial location. We first note that overall decoder accuracy starts reasonably low near the beginning of the trial indicating reduced phase-locked responses to the male speaker (from either left or right side) very early in the scene. Throughout the trial, when the probability of speaker convergence is low (i.e., greater spatial separation), the decoder accuracy yields a high performance near 80\%. This performance drops sharply between 3-4.2 seconds to around 37\% when there is a high probability of one speaker moving towards the side where another speaker already is, leading to less spatial separation between speakers. Spearman rank correlation between decoding accuracy and the speaker convergence probability distribution reveal a strong negative correlation ($rs = -0.684$), indicating that accuracy tends to decrease as the speakers converge and spatial separation between speakers decreases.  A permutation test (1,000 iterations) confirms that this relationship is statistically significant (p = 0.001).

For the Spatial condition (right), the histogram probability represents the likelihood of a speaker starting to switch from the attended (right) side of the auditory scene (attended-side departure).  This distribution is most relevant to the spatial mode as it reflects disruption in spatial stability of the scene. At the beginning of the trial, the decoder accuracy starts reasonably high given that participants are immediately cued to the direction to attend to. This cueing likely underlies increased encoding of all voices in the attended direction though might still require segregation of these voices to facilitate target detection as seen in the buildup effects of Figure \ref{fig:temproal_dynamics}A. The accuracy starts dropping around 2.2. seconds as the probability a speaker leaves the attended side increases.  The decoder accuracy reaches its lowest point, below 20\%, just before 4 seconds, when the attended-side departure probability is highest.  From 4-4.5 seconds, as the probability decreases, the accuracy increases.  A second peak in switch probability occurs around 4.5-5.5 seconds which again causes a dip in decoder accuracy.  Again, Spearman rank correlation between the decoding accuracy and the attended-side departure probability distribution reveals a strong negative correlation ($rs = -0.302$), indicating that accuracy tends to decrease as the probability that a speaker switches away from the attended side increases.  A permutation test (1,000 iterations) confirms that this relationship is statistically significant (p=0.037).

\subsection{Global attention operates using a source sampling strategy by examining different voices as the scene unfolds}

In contrast with the selective attention modes where an attended and unattended stream emerges, the global attention task requires subjects to attend to the scene as a whole. To gain insight into neural mechanisms underlying this task, we examine decoders trained on each speaker in the scene (male, female 1, female 2) and each side (right, left) for each of the 3 attentional modes. For each analysis window, the decoder with the highest correlation is denoted as winner.

\begin{figure*}[t]
    \centering
    \includegraphics[width=0.85\textwidth]
    {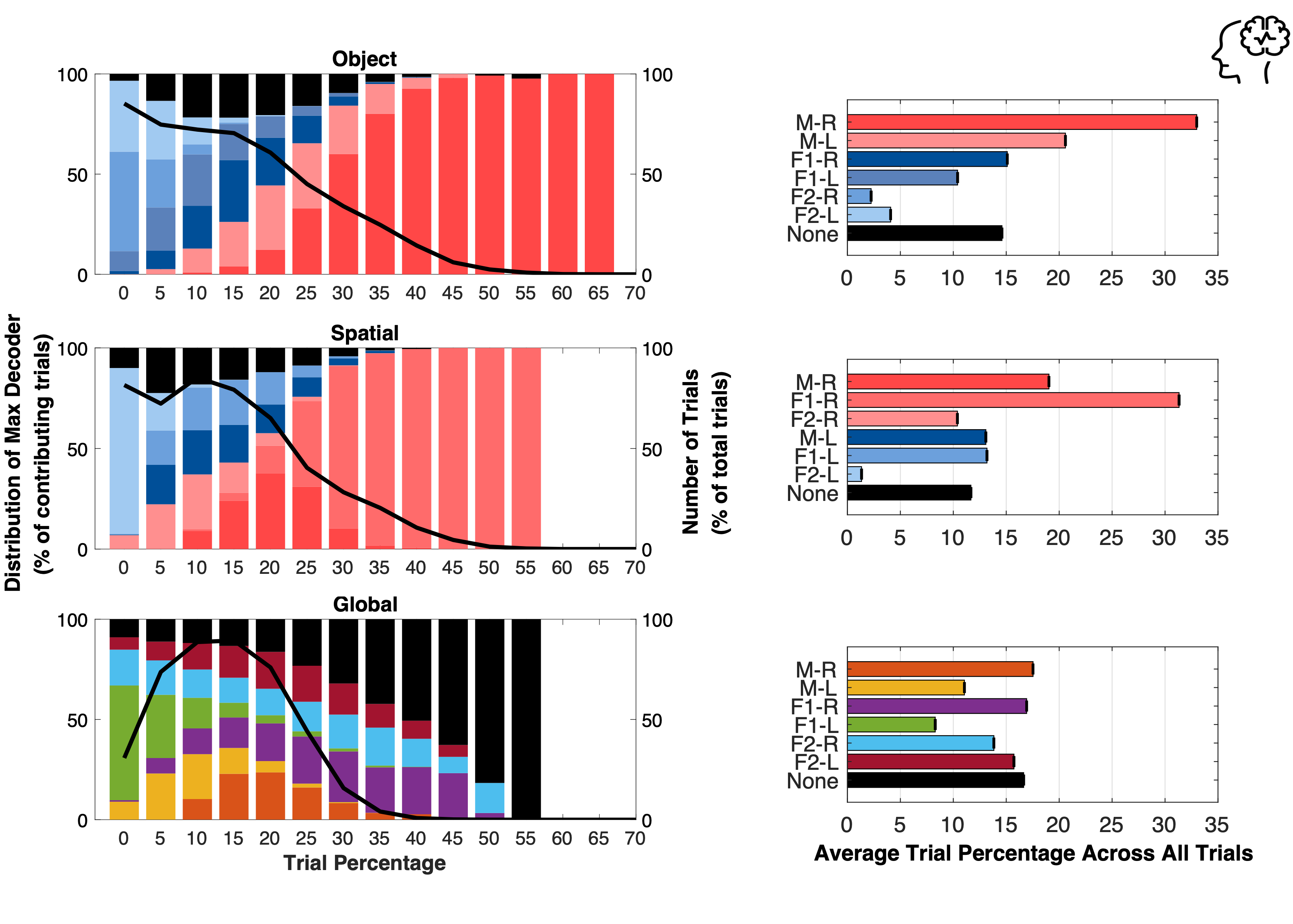}
    \caption{\textbf{Left Panel}: Distribution of trial percentages during which each decoder is dominant.  \textbf{Right panel}: Average trial percentage for each decoder across all trials.  \textbf{Both panels}: \textit{Top}: Object, \textit{Middle}: Spatial, \textit{Bottom}: Global.  For object and spatial plots, red represents decoders in the attended stream and blue represents decoders in the unattended stream.  For all plots, black represents no dominant decoder.  M-R: male-right, M-L: male-left, F1-R: female1-right, F1-L: female1-left, F2-R: female2-right, F2-L: female2-left}
    \label{fig:decoder_occupation}
\end{figure*}

Figure \ref{fig:decoder_occupation} (Left Panel) shows the distribution of trial time percentages during which each decoder is considered the dominant decoder.  A decoder is determined to be the dominant decoder at a given time point if it has the maximum value among all decoders, exceeds a predefined threshold, and surpasses the next-highest decoder by a minimum margin (see \ref{section:decoder}).  The x-axis (Trial Percentage) represents the percentage of a trial where a given decoder is the dominant decoder.  For example, if a trial is 8 seconds, x = 25 means 25\% of the trial, or 2 seconds of the trial.  The right y-axis (Number of Trials) represents the percentage of the total number of trials that actively contribute to each time point (black line).  For example, in the object condition, at x = 25, 44\% of all trials have at least one decoder that is dominant for 25\% of the trial.  The left y-axis (Distribution of Max Decoder) represents the proportion of trials (as a percent of the contributing trials, right y-axis) in which each decoder dominates at each point in time.   For example, in the object condition, at x = 25, 44\% of all trials contribute, and the dominant decoder is the Male-Right for 33\% of the contributing trials, Male-Left for 32\% of the contributing trials, Female1-Right for 14\% of the contributing trials, Female2-Right for 6\% of the contributing trials, and the dominant decoder could not be determined for the remaining 15\% of the time; the remaining two decoders (Female1-Left, Female2-Left) are not the dominant decoders in any trial for this amount of time.  Figure \ref{fig:decoder_occupation} (Right Panel) shows the average trial time percentage for each decoder across all trials. For the object and spatial modes, the shades of red indicate  decoders in the attended stream, and the shades of blue indicate decoders in the unattended stream.  Because there are no attended and unattended decoders in the global mode, each decoder is assigned a different color.  In all three modes, the black indicates that a dominant decoder could not be determined.

In the object and spatial modes, the subjects' attention tends to be on the attended decoders (the two male decoders for the object mode, and the three right decoders for the spatial mode) more often than the unattended decoders. This is clearly indicated by the prominence of the red-shaded decoders in the top and middle panels. In the case of global attention, an interesting phenomenon emerges. Figure \ref{fig:decoder_occupation}, bottom row shows that global attention tends to switch among the six decoders, implying that subjects tend to sample each sound source in the auditory scene, with no one voice (or decoder) occupying more than 18\% of any given trial and most decoders being dominant between 8\% and 17\% of each trial. This uniform distribution supports a source-sampling pattern for the global model of attention suggesting that participants frequently sample different voices in the scene. 

\subsection{Global attention benefits most from bottom-up attention compared to selective modes}
\begin{figure*}
    \centering
    \includegraphics[width=\textwidth]{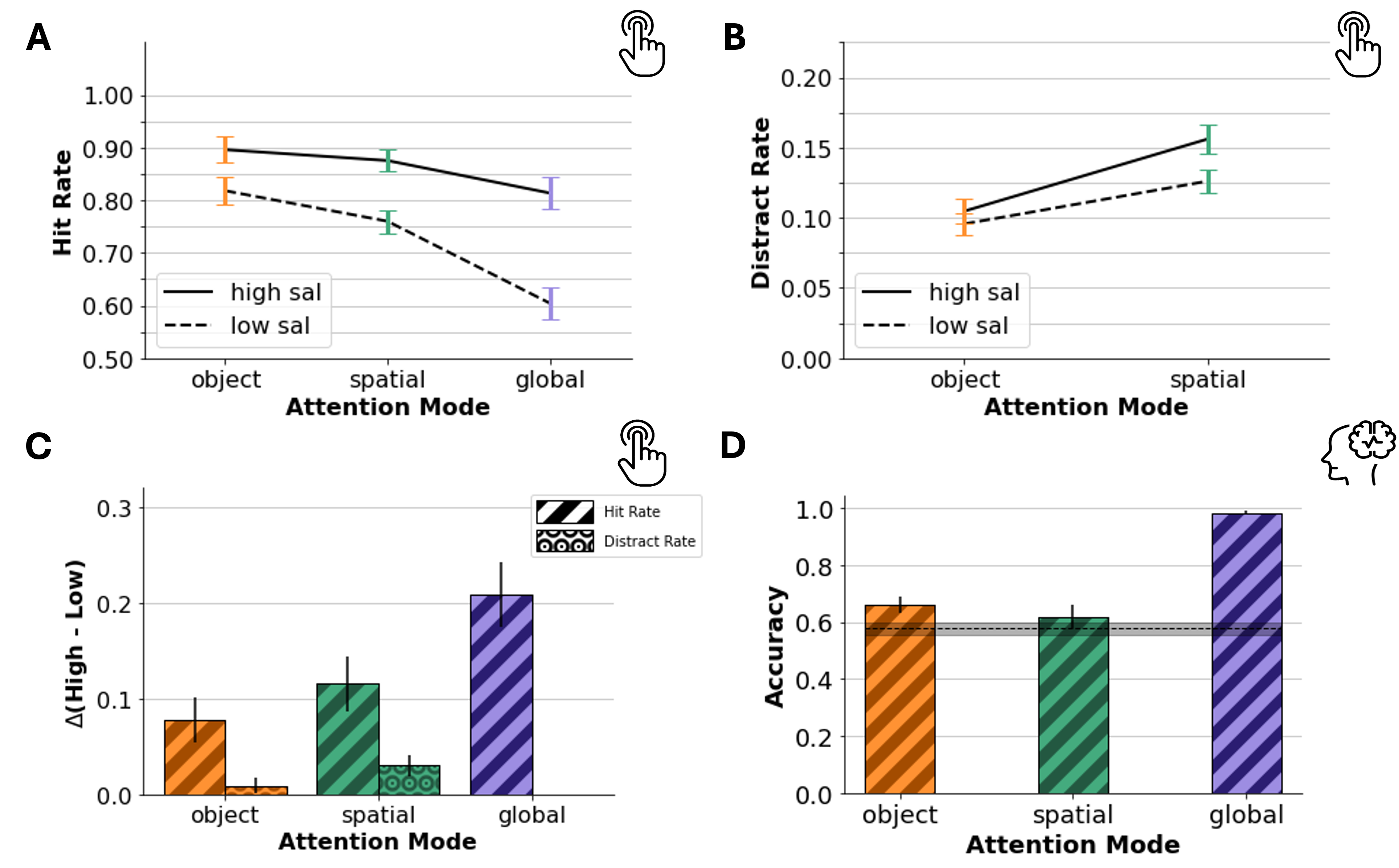}
    \caption{\textbf{(A)} Hit rates for high and low salience targets under each mode of attention.  \textbf{(B)} Distract rates for high and low targets under object and spatial modes of attention. \textbf{(C)} The greatest difference in hit rate occurs for the global mode, while the greatest difference in distract rate occurs for the spatial mode. \textbf{(D)} Neural signature classification of high and low salience targets for each attention mode.  Similar to the behavioral responses, classification accuracy is greatest for the global mode.}
    \label{fig:hi_lo_sal}
\end{figure*}

Dividing the targets into high and low salience events reveals that global attention shows the most gain from the salience level of the target compared to selective attention. Figure \ref{fig:hi_lo_sal}A shows hit rates for high versus low salience targets for each mode of attention.  Selective modes of attention strongly boost perception of low-salience targets compared to the global mode of attention, while high-salience targets are detected to a similar degree under all modes of attention. The hit rate in the global mode for low salience targets is only around 60\%, but increases significantly for high salience targets around 82\%. 

This nonlinear behavior of target salience under different attention modes is confirmed with a two-way repeated-measures ANOVA on hit rates across attention mode and salience state and reveals that there is a statistically significant interaction between attention mode and salience state (F=6.57, p=0.002).  Simple main effects show that both attention modes and salience states have a statistically significant effect on the hit rate (F=25.42, p=7.92e-10 for attention mode and F=53.71, p=9.90e-10 for salience state). A Friedman test conducted across attention modes for each salience level indicates significant effects of attention mode for each salience level (high salience: $\chi^{2}$(2) = 10.65, p = 0.005; low salience: $\chi^{2}$(2)=32.04, p=1.10e-7).  Bonferroni-corrected Wilcoxon signed-rank tests are then used to examine pairwise differences between attention modes for each salience level.  For high salience targets, hit rates do not differ significantly between object and spatial (W=198.0, p=1) or between spatial and global (W=130.5 p=0.067), but object and global show a significant difference (W=97.5, p=0.028).  For low salience targets, all pairwise comparisons are significant: object and spatial (W=228.5, p=0.042), object and global (W=46.5, p=1.85e-7), and  spatial and global (W=113.0, p=6.54e-5).  Within each attention mode, Wilcoxon signed-rank tests comparing high- and low-salience targets confirm significantly higher hit rates for high-salience targets across all attention modes (object: W=198.5, p=2.24e-02; spatial: W=152.5, p=1.11e-04; global: W=183.0, p=4.33e-06, Bonferroni-corrected)

Looking at the distract trials (Figure \ref{fig:hi_lo_sal}B), false positive distract rates are slightly higher for high salience targets compared to low salience targets, though this trend is more evident in the spatial mode than the object mode.  A two-way repeated-measures ANOVA across attention mode and salience state on the false positive distract rate reveals a statistically significant interaction between attention mode and salience state (F=5.63, p=0.020), suggesting the effect of salience state on distract rate varied across attention modes.  Simple main effects show that attention mode has a significant effect on the distract rate (F=20.8919, p=6.8924e-6), implying distract rates differ for each task, with object having a lower distract rate.  Similarly, salience state also has a significant effect on the distract rate (F=4.5829, p=0.0330), with lower distraction rates for low salience targets. Wilcoxon signed-rank tests for each mode indicate that there is no significant difference between high- and low-salience distractors for the object mode (W=797.0, p=0.384), but there is for the spatial mode (W=552.5, p=0.008).  Furthermore, Wilcoxon signed-rank tests for each salience level indicate a significant difference between attention modes for both high-salience (W=318.0, p=6.51e-6) and low-salience (W=438.0, p0.001) distractors.  Figure \ref{fig:hi_lo_sal}C shows the difference between high and low salience target hit rates for each attention mode, as well as the difference between high and low salience target distract rates for each attention mode and captures the larger difference in hit rate observed in the global mode. 

Using the neural responses to classify the salience of each target as high or low saleince reveals the same benefit of high salience targets in the global attention.  Figure \ref{fig:hi_lo_sal}D shows the results of high/low salience classification  from the neural data of correct trials.  The greatest neural signature discrimination between high and low salience targets occurs for the global mode (98.3 \textpm 2.7\%) compared to the object mode (66.0 \textpm 1.2\%) and spatial mode (61.9 \textpm 1.2\%), mirroring the trend observed in the behavioral data depicted in Figure \ref{fig:hi_lo_sal}C.  Interestingly, object and spatial attention performance is only slightly though still significantly above a baseline chance classifier (57.7 \textpm \ 2.35\%).  This suggests that salience may not be sufficiently distinguishable in object and spatial attention but plays a significant role in driving performance in global attention.

\section{Discussion}

Auditory attention allows listeners to make sense of real-world environments by selectively enhancing relevant sounds while suppressing irrelevant ones. This capacity has long been studied through paradigms such as the “cocktail-party” effect, where listeners are able to track a voice of interest despite simultaneous competing sources \cite{Cherry1953,Haykin2005}. Previous research has shown that attention can be directed toward specific auditory features such as spatial location \cite{Mondor1995,Ahveninen2006}, toward coherent auditory objects like speech streams \cite{Best2008,Ding2012}, or more diffusely in global or free-listening modes that lack task-driven focus \cite{Naatanen2007}. However, whether these modes reflect distinct neural mechanisms or a unified attentional system operating at different levels remains a key unresolved question in auditory neuroscience.

The present study directly addresses this issue by contrasting object-based, feature-based (spatial), and global attention while holding the acoustic input constant. This allows us to isolate attentional strategy from stimulus-driven factors. Empirical findings indicate that these attentional modes are not only behaviorally distinct but also engage separable neural mechanisms, supporting the idea that auditory attention is implemented by multiple, functionally dissociable systems. These findings extend and clarify a range of earlier observations from the neuroimaging literature on spatial tuning and speech tracking \cite{Altmann2008,Degerman2008,Petkov2004,Fritz2003NatNeur} to models of attention that posit feature and object selection as coexisting yet distinct processes \cite{kravitz2011,Mayer2012} .

In line with our first hypothesis (H1), object-based attention yielded the strongest behavioral performance, with significantly higher hit rates and lower distraction compared to spatial attention. These findings replicate and reinforce earlier work showing that attention directed toward auditory objects results in enhanced perception and cortical tracking \cite{Kerlin10a,Mesgarani2012}. What is notable from the findings is that these benefits were observed despite the fixed acoustic input hence ruling out confounds related to acoustic variability or task difficulty and confirming that the attentional mode itself shapes perception and behavior. EEG analyses revealed distinct neural signatures for each mode: Object attention produced bilateral activation in temporal areas, while spatial attention elicited the characteristic parietal alpha lateralization associated with spatial focus \cite{Banerjee2011OscillatoryMechanisms,Frey2014SelectiveTask}. These results suggest that even when feature-based attention is effective, object-based tracking provides more robust engagement, particularly when the auditory scene is dynamic and sources move in space. Yet, object-based attention did not always outperform feature-based attention. In the case of salient targets, spatial and object tasks yielded statistically similar detection rates. This results argues that when bottom-up cues are strong and diagnostic, feature-based selection can match the effectiveness of object tracking. Such flexibility aligns with visual attention models showing that strong feature salience can drive efficient selection even in the absence of object continuity \cite{Frey2014SelectiveTask,Wegener2014}. However, under lower-salience conditions, the difference between the two modes was more pronounced, suggesting that object attention confers resilience when perceptual demands are high or cues are ambiguous. This hierarchical relationship mirrors the idea proposed in auditory scene analysis that proto-objects may form from early feature grouping and serve as the substrate for object-based attention \cite{Shinn-Cunningham2008,Bregman1990}.

It is worth noting that although in this work we have only tested feature-based attention to acoustic locations, neuroimaging results demonstrating attentional modulation for a variety of features suggest that similar effects could possibly be seen for directed attention to other features. This is especially the case considering that there does not appear to be feature-dependent differences in feature-based attention in vision, and that visual and auditory attention seem to share many common mechanisms \cite{Scholl2001,Busse2005}. One distinction from vision, however, is in the treatment of the spatial dimension. Space-based attention has traditionally been treated as a separate form of selective attention in vision \cite{soto2004,kravitz2011}, though it has been suggested that it could be unified under the same framework as feature-based attention \cite{Maunsell2006}. However, there is little evidence supporting space as a special feature in audition. Studies that have investigated the effect of attention to frequency and space have suggested the two features operate under the same fundamental process \cite{Krumbholz2007,Maddox2012}. It is also worth considering that space in audition is derived from neural computations on signals reaching the two ears, in a similar manner to pitch or other acoustic features. Even if spatial attention differed significantly from feature-based attention, the current results still demonstrate that auditory attention can operate in distinct modes, with spatial or feature-based attention differing from global and object-based attention.

In contrast to the selective modes, global attention served as a baseline in our paradigm and emerged not as passive monitoring but as an active, dynamic strategy. Decoding of stimulus envelopes revealed that listeners attending globally did not maintain a combined percept of the scene as a whole but rather cycled through individual sources. The analysis of decoding performance supported the hypothesis H2 that no decoder dominated the trial and the system appeared to sample streams serially. This “source-sampling” behavior provides new empirical support for theoretical claims that global attention involves exploratory scanning guided by salience and expectation \cite{Oliva2007,Kaya2014}. These findings challenge the notion that global attention reflects a degraded or noisy version of selective listening. Instead, it appears to be a qualitatively distinct attentional state, particularly suited for detecting unexpected events across a complex scene.

This interpretation is further supported by the temporal dynamics of behavior and EEG responses. Following early spatial shifts in speaker location, performance in all attentional tasks initially declined but quickly recovered, particularly for object attention. These build-up effects were limited to early shifts. Later in the trial, attention had stabilized, and further spatial changes no longer disrupted performance. These findings align with previous work showing that object formation takes time but is resilient to spatial perturbation once it is established \cite{Cusack2004,Micheyl2005}. Importantly, we interpret these findings not as evidence of object formation per se, but of rapid object updating or re-binding within a stable perceptual framework. The transient drop in accuracy after early shifts may reflect reorientation within an established object map rather than full re-segregation.

While the dynamics of selective and global attention were largely shaped by task goals, the influence of bottom-up salience introduced a different dimension and revealed that the relationship between bottom-up salience and attention mode was highly nonlinear. In support of hypothesis H3, high-salience targets significantly improved detection in the global task, but had much smaller effects under selective attention. This suggests that once attention is allocated to a particular object or spatial stream, the system is less influenced by salience in competing sources. This pattern is consistent with studies showing suppression of unattended salient stimuli during focused tasks \cite{Wegener2008,Taya2009,Freeman2014}. Our neural classification of salience echoed this behavioral pattern: high- vs. low-salience targets were most discriminable in global attention (98\%), but only marginally so in object and spatial tasks (${\sim}62\%$). These findings add to a growing body of literature suggesting that bottom-up salience is modulated by top-down task constraints, and further suggest that the influence of salience is greatest when attentional goals are weakly specified or diffuse.

Taken together, these findings clarify long-standing ambiguities in the auditory attention literature and offer a framework for understanding how attentional systems interact in real-world listening. By testing object, feature, and global attention in a tightly controlled yet naturalistic context, we provide evidence for distinct processing modes with different temporal dynamics, neural signatures, and sensitivity to salience. These results unify previous psychoacoustic, imaging, and electrophysiological findings and suggest that attentional mechanisms are flexibly deployed based on task demands and scene statistics. Future work should investigate how these modes interact and transition, whether they are instantiated by overlapping or independent cortical networks, and how they contribute to auditory perception in more complex, multi-modal environments.

\section{Materials and Methods}
\label{sec:Materials}

\subsection{Experimental Paradigm} 
Participants in both behavioral and neural experiments were presented with sound mixtures consisting of simultaneously played German audiobook narration extracts by three voices (a young adult female, an old adult female, and a young adult male). The three voices moved spatially throughout the duration of the trial, following a pattern explained next. In test trials, a pitch shift of one of the voices was introduced at some point during the trial. The experiments engaged the participants in three consecutive tasks divided into three experimental blocks:

\begin{enumerate}
\item In the first block (global attention), participants were asked to attend to the sound mixtures and indicate whether \emph{any} of the voices had a change in pitch. Behavioral responses were recorded at the end of each trial as two-alternative forced-choice Yes/No response in both behavioral and neural experiments.

\item In the second block (spatial attention), participants were instructed to attend only to voices in the right ear and flag any change in pitch. Detection of pitch changes in voices on the left would be considered a false alarm.

\item In the final block (object attention), participants were instructed to pay attention to the male voice regardless of its spatial location and flag the presence of pitch changes, while ignoring the female voices. 
\end{enumerate}

All participants performed the global condition as their first block and then were randomly assigned the spatial and object block in a counter-balanced fashion. Critically, the \emph{same} 50 trials were presented in all three blocks (with trial order randomized in each block); therefore allowing careful control over acoustic parameters across attentional conditions. Participants performed a short training block before the experiment to ensure they understood the global task. Subjects had no prior knowledge of the spatial/object tasks until the end of the global block. The choice of German narrations was deliberate, as participants in both behavioral and neural experiments did not speak any German, hence relying primarily on the acoustic characteristics of the scene without any linguistic or semantic processing. Both behavioral and neural experiments took place inside a soundproof booth and stimuli were delivered with headphones (Sennheiser HD595).

\subsubsection{Stimulus design}
The 3 voices were manually extracted from public domain Librivox German book narrations (Male: \url{https://librivox.org/ein-vade-mecum-fur-den-hrn-sam-gotth-lange-pastor-in-laublingen-by-gotthold-ephraim-lessing/}, Female 1: \url{https://librivox.org/menschen-im-krieg-by-andreas-latzko/}, Female 2: \url{https://librivox.org/das-letzte-maerchen-by-paul-keller/}) recorded at 22050 Hz.  From these narrations, 81 male segments were extracted from chapters 2 and 5, 61 female 1 segments were extracted from chapter 1, and 58 female 2 segments were extracted from chapter 10.  The overall pitch range of the speakers was approximately A3-D4.  Segments were chosen to have a prosody to sound like spoken single sentences, with no regard to meaning of the words spoken or whether the segment contains actual sentences.  Listeners need continuous speech to follow speakers in a busy auditory scene, so segments were manually processed to reduce silent periods, including narrator pauses and words spoken quietly. The 50 trials were constructed by selecting one unique segment for each narrator such that the length difference between the three segments would be less than 300ms.  

The task was to detect a segment with a change in pitch. Fifty trials with unique sentences were constructed, 10 of which contained no pitch altered segments (control trials). The target specification in the remaining trials was as follows: 10 male-right, 15 male-left, 10 female-right, 5 female-left. The target was a 1 s long segment of the speech that was altered to have slightly higher (if female) or lower (if male) pitch. The right target segments were in the absolute right for the duration of the target, but the left targets could be anywhere between center and absolute left. 

\subsubsection{Target construction}
Targets were 1 second long segments of voice being manipulated to have a modified pitch.  Pitch manipulation was performed by time dilation for male targets and compression for female targets with a phase vocoder.  Targets in the male voice were 3 semitones lower than the natural speech, while targets in the female 1 voice were 3 semitones higher.  Male targets never appeared before the first direction change of the male from right to left. Female 1 targets could appear on the left side before the direction change. There were no female 2 targets. Target segment onsets and offsets, as well as speech immediately prior to and following the target were smoothed by a 10ms ramp to avoid abrupt transitions in sound. 

\subsubsection{Spatial parameters:} 
Each trial consisted of the 3 voices narrations played simultaneously over headphones. Each speaker started at a specific spatial location at the start of every trial: The male always started on the right, female 1 always started on the left, and the female 2 always started in the center.  After a brief period, the speakers start switching locations in space.  Once one voice reached the new location, the speaker remained there for a few seconds while the voice that was originally at that location moved to the opposite side.  This movement pattern happened 1-3 times in each trial.

In order not to bias the audio towards any one direction, the voice movements were constructed such that there would be no gaps of speech at absolute right or absolute left at any point during the trial.  Voices never overlapped in any direction except for the brief moments when one voice was reaching a direction just as another speaker was leaving.  Additionally, no voice lingered in the center after stimulus onset: Voices were either moving between right and left spatial locations or entirely at the right or left sides.  Stimulus length varied between approximately 5-10 seconds.  The speed at which a voice moved between the right and left directions was constant, with a voice taking 1.2 seconds to move from one side to the other.  The direction shifts started at a random time after the first second, and the number of shifts in each trial depended on trial length.

To simulate voice positions in 3-D space, a head-related transfer function (HRTF) was applied.  The HRTF was recorded on a mannequin (Neumann KU 100) under the same conditions in which human HRTFs are recorded.  The NH172 HRTF was used from the ARI HRTF database.  Trajectories for each voice were constructed between -90 degrees (left) and 90 degrees (right) denoting their position at each point in time.  Spatial dynamics of the scenes were increased by adding jitter to the trajectory when speakers had stable position at left, center or right.  That is, instead of a straight trajectory, a sinusoid with a period of 5 spanning 50 degrees was inserted.  At the time of the direction switch, the trajectory moved from the middle of the left or right jitter (65 or -65 degrees) in a linear line lasting 1.2 seconds.  The female 1 and male voices were permitted to move between the left and right sides, while female 2 only moved between the right side and center.  Female 2 moved at the same speed as the other two speakers, but the movements took only as long as necessary to fill in gaps in the absolute right.

The spatial treatment of the voices resulted in 3 stereo signals: male-right, male-left, female1-right, female1-left, female2-right, female2-left. These tracks were combined together resulting in a rich 3 D scene that was presented to participants.

\subsection{Participants}

All behavioral and neural experimental procedures were approved by the Johns Hopkins Homewood IRB; and all participants gave informed consent and were compensated. 90 subjects (55\% females, 18-31 years of age) participated in the behavioral experiments. 18 subjects had a negative Dprime and were removed from further analysis due to inability to perform the task. 24 subjects (58\% females, 18-31 years of age) participated in the neural experiments. The same 50 trials used in the behavioral experiment were presented to the subjects while neural data was collected via EEG.

\subsection{EEG acquisition and preprocessing}
\label{eeg}
EEG measurements were obtained using a 64-channel Biosemi Active Two system.  EEG data were analyzed using MATLAB (Mathworks Inc, MA), with FieldTrip \cite{Oostenveld2011} and Noisetools analysis tools and Python toolboxes and custom scripts.  Neural signals were first down-sampled to 256 Hz, then demeaned.  Next, bad channels were identified and interpolated using neighboring good channels.  The data were detrended, and any remaining bad channels were again identified and interpolated.  Eyeblinks were eliminated using time-shifted PCA, and outliers were detected based on a weighted correlation structure and removed.  Finally, the data were re-referenced by subtracting a weighted mean.

\subsection{Data analysis}

\subsubsection{Behavioral responses}
Correct responses (i.e. detect targets in test trials) and false alarms were obtained for each attentional mode and each subject. All results were corrected for multiple comparisons using Holm-Bonferroni correction to confirm statistical significance \cite{Snedecor1989} and results post correction were reported. Residuals were checked for normality using the Shapiro–Wilk test ($p < 0.05$).  All analyses were repeated whether the spatial or object task was performed first and revealed no differences due to the order of the experiment.

\subsubsection{Behavioral responses in behavioral only vs. behavioral and neural experiments}

Behavioral responses were obtained from two datasets: one collected in a behavioral-only experiment with a larger sample (90 subjects), and one collected concurrently with neural recordings (24 subjects).  To assess whether differences in sample size influenced the results, permutation tests were conducted for hit rate, false positive rate, and d-prime.  For each metric, 24 subjects were randomly sampled without replacement from the larger dataset 1,000 times, and the mean was computed for each subsample.  The resulting distribution of sub-sampled means was then compared to the mean of the smaller dataset.  p-values were calculated as the proportion of sub-sampled means that were as or more extreme than the observed mean in the smaller dataset.

\subsubsection{Salience calculation}
\label{stimulus_salience}
Target salience was computed using a bottom-up attention model \cite{Kaya2014} trained on human ratings of natural salience scenes.  The model builds statistical predictions among a variety of acoustic features.  Salience for each feature is then derived as a function of the deviation from the predicted feature value at each point in time.  To reduce noise, only the interactions for the maximum spikes along each feature at the time of the target were calculated, resulting in one feature vector per trial.  To classify salience levels, subject responses were used as ground-truth data.  Trials were assigned 0 if less than half the subjects heard the target, and 1 if half or more of the subjects heard the target.  Finally, each trial was classified as low or high salience depending on whether salience predictions from logistic regression were less than 0.5, or greater than or equal to 0.5, respectively. 

\subsubsection{Neural heatmaps}

Heatmaps were generated from the EEG data processed as outlined in Section \ref{eeg}.  First, the EEG data for each trial was normalized to zero-mean and unit-variance.  Next, the portions of the trial corresponding to the non-switch regions were selected for analysis.  This normalization and selection process was repeated for all subjects and trials.  Next, a sampling procedure was used in which all trials from approximately two-thirds of the subjects were selected and averaged together.  This process was repeated 1,000 times to generate 1,000 representative sample trials.  Finally, Python's MNE toolbox was used to visualize the heatmaps.  Channel significance was determined via bootstrap confidence intervals on the subject means.  Subject-level means were first obtained by averaging across trials within each subject.  For each channel, a 95\% bootstrap confidence interval (1,000 resamples) was computed across subjects.  Channels whose confidence intervals did not include zero were considered significant.

Analysis of oscillatory activity across scalp channels used 5 anatomically distinct channel clusters following the grouping used in 
https://www.eneuro.org/content/12/2/ENEURO.0275-24.2025.full
The analysis focused on ``frontal'',``bilateral temporal'', ``central'',``parietal'' and ``occipital'' groups. For each channel cluster, the average power spectrogram was derived using a 100msec window with 50\% overlap. Power in specific frequency bands was then derived over the theta (4-7Hz), alpha (8-12Hz), beta (12-20Hz) and gamma (30-125Hz). For each band and channel cluster, a one-way repeated-measures ANOVA was conducted to examine the effect of attention mode on EEG activity. Post-hoc pairwise comparisons (using Bonferroni-corrected t-tests) were applied to explore individual differences between attention modes. Given the limited number of trials in each condition across bands, channels and subjects, other statistical tests using bootstrapping were also performed to confirm observed trends. Effects reported here were confirmed using both methods.

\subsubsection{Neural classification of attentional modes}
\label{classifier}

A logistic regression classifier was trained on EEG data to classify the data as belonging to one of the attention modes: global, object, spatial (Figure \ref{fig:attention_modes}D, \ref{fig:attention_modes}E, and \ref{fig:temproal_dynamics}D) or to classify responses as corresponding to a high or low salience target within each mode (Figure \ref{fig:hi_lo_sal}D).  The EEG was processed as outlined in Section \ref{eeg}.  Next, Multiway Cannonical Correlation Analysis (MCCA) \cite{deCheveigne2018b} was applied to the EEG data, reducing the original 64 channels to 30 canonical components.  These components were then encoded into five frequency bands ($\delta$=1-2, $\theta$=3-7, $\alpha$=8-15, $\beta$=16-30, $\gamma$=31-100 Hz). Finally, the average power within each band was computed to serve as the input features to the classifier.  To evaluate classifier performance and assess stability, the model was trained using a grid search over regularization parameters, specifically, a range of values for the inverse regularization strength (C) and the elastic net mixing parameter (L1 ratio) using 5-fold cross-validation for model selection.   To ensure robustness to data partitioning, this entire procedure was repeated ten times, each with a unique random seed controlling the stratified shuffling.  This resulted in 10 independent classifier performance estimates, from which the mean and standard deviation were computed to assess overall classification accuracy and variability.

To evaluate classifier performance relative to chance, a null model was established using label-shuffled classification.  For pairwise classification between groups, separate null models were computed for each of the three attention mode pairs and averaged to obtain a single baseline value (Figure \ref{fig:attention_modes}D).  For high/low salience level classification within each attention mode, null models were computed independently for each mode and similarly averaged (Figure \ref{fig:hi_lo_sal}D).  For the three-way classification, a single null model was computed across all classes (Figure \ref{fig:attention_modes}E).  These null models were used to quantify and compare classifier accuracy relative to chance.

\subsubsection{Neural decoding}
\label{section:decoder}

The stimulus envelope was reconstructed from neural recordings, in order to evaluate the tracking of speech envelopes under different attentional conditions  \cite{Aiken2008, Ding2012}. The neural responses were processed according to the procedure in Section \ref{eeg} then Python's MNE toolbox was used to fit decoding models to estimate the stimulus (envelope) from neural responses.

In this experiment, each trial had three voices (male, female 1, female 2) and two sides (right, left), for a total of six acoustic signals per trial.  For each attentional mode (object, spatial, global), subject-specific decoders were estimated for each acoustic signal (male right, male left, female 1 right, female 1 left, female 2 right, female 2 left).  A leave-one-out cross-validation was performed to find the optimal regularization parameter from a set of values ranging from $10^{-4}$ to $10^3$ and optimal time lags ranging from $0$ to $500ms$.  To train the model, all trials but one were concatenated then normalized to zero-mean and unit-variance , while the remaining trial was normalized and reserved for testing.

This procedure resulted in 6 predicted stimulus envelopes for a given test trial. Pearson's correlation between predicted and actual envelopes was computed over 1-second windows, with varying resolutions (between 50\% and 90\% overlap depending on analysis). The analysis was performed for each attentional mode, subject, and trial stimulus, and quantitatively similar results were obtained for different window sizes and overlap values.

Correlation values over time were organized in three $N$ x $T$ x $D$ matrices (one for each attention mode) where $N$ is the number of trials in the experiment, $T$ is the time index sampled over different windows and $D$ is the number of decoders (6 corresponding to 3 stereo stimuli). For each task, the mean across trials resulted in a correlation value for each window and each decoder. Mean correlations across decoders were compared in two ways:  combining decoders into attended (foreground) and ignored (background) streams based on attentional condition; or computing the maximum correlation across all 6 decoders.

\noindent \emph{Attended vs. Ignored decoders:}

The attended/ignored decoder grouping was performed as follows. For the object task, the attended stream ($A_{obj}$) and unattended stream ($U_{obj}$) decoders were defined by combining together decoder outputs of the male voice (attended stream) and combining decoder outputs of the female1 and female2 voices (ignored streams), following the rule:

\begin{eqnarray*}
    A_{obj} & = & \max \left( \text{male-left, male-right} \right) \\
    U_{obj} & = & \max \left( \text{female1-left, female1-right, female2-left, female2-right} \right)
\end{eqnarray*}

For the spatial task,  the attended stream decoder ($A_{spa}$) was defined over all voices on the right channel, while the 
 unattended stream decoder ($U_{spa}$) was defined over all voices on the left channel, following the rule:
 
 \begin{eqnarray*}
    A_{spa} & = & \max \left( \text{male-right, female1-right, female2-right} \right) \\
    U_{spa} & = & \max \left( \text{male-left, female1-left, female2-left} \right)
\end{eqnarray*}

No such grouping was feasible for the global condition. For the object and spatial tasks, the maximum correlation was obtained for the attended and ignored streams, producing a $1\ \textnormal{x}\ T$ vector of correlation values. 

To analyze decoding accuracy, the classification rules described above were applied within a bootstrap sampling framework to reduce the impact of noise and account for experimental complexity and variability.  For each iteration (1,000 total), the trials were randomly sampled with replacement and averaged to form a representative trial.  The attended and unattended groups were defined according to the specified rules, and each resampled trial was flagged as correct (1) if the attended output exceeded the unattended output, or incorrect (0), otherwise.  Accuracy and variability were then computed by averaging across all iterations for each time window.

To evaluate the relationship between the accuracy and the probability distribution, Spearman rank correlation coefficient, a non-parametric measure of monotonic association that does not assume normality, was computed.  Because both the accuracy and probability distribution failed a normality test (Kolmogorov–Smirnov, $p < 0.05$), we assessed statistical significance using a permutation test rather than relying on standard correlation p-values.  Specifically, we computed the observed Spearman correlation, then generated a null distribution by randomly permuting one variable 10,000 times and recalculating the correlation at each iteration.  The p-value was computed as the proportion of permuted correlations that were as or more extreme than the observed value (two-tailed test).  This procedure allowed us to determine whether the observed correlation was significantly different from chance.

\noindent \emph{Decoder switching:}

A second analysis looked at all 6 decoders as a function of time across the different attentional tasks and evaluated the \emph{duration} in which each of the decoders was the highest across each trial. The analysis computed the percentage of a trial each of the 6 decoders was maximum. 30\% of trials were discarded from the analysis as they contained outlier counts for at least one decoder.  Outliers were defined as values 1.5 times the inter-quartile range greater than the third quartile or less than the first quartile.  To further mitigate noise in the data arising from the complex experiment paradigm, a bootstrap sampling method was implemented in which a subset of the trials from each attention mode were randomly selected and averaged together to form a representative trial for each mode; this process was repeated 1,000 times to get a collection of representative trials.  For each window in the trial, the decoder with the highest correlation was deemed to be the decoder to which the subject was attending at that point in time.  If the top two correlations were within a certain threshold (0.2\% of each other), or if the maximum correlation was below a certain threshold (0.01), no decoder was selected as the attended decoder.  For each of these new sampled trials, once again the number of windows that each decoder was the maximum decoder were counted.

\section{Acknowledgments}
This work was supported in part by ONR N00014-23-1-2050 and NSF 2444353-01.

\bibliographystyle{IEEEtran_nourl}
\bibliography{references}

\begin{thebibliography}{10}
\providecommand{\url}[1]{#1}
\csname url@samestyle\endcsname
\providecommand{\newblock}{\relax}
\providecommand{\bibinfo}[2]{#2}
\providecommand{\BIBentrySTDinterwordspacing}{\spaceskip=0pt\relax}
\providecommand{\BIBentryALTinterwordstretchfactor}{4}
\providecommand{\BIBentryALTinterwordspacing}{\spaceskip=\fontdimen2\font plus
\BIBentryALTinterwordstretchfactor\fontdimen3\font minus \fontdimen4\font\relax}
\providecommand{\BIBforeignlanguage}[2]{{%
\expandafter\ifx\csname l@#1\endcsname\relax
\typeout{** WARNING: IEEEtran.bst: No hyphenation pattern has been}%
\typeout{** loaded for the language `#1'. Using the pattern for}%
\typeout{** the default language instead.}%
\else
\language=\csname l@#1\endcsname
\fi
#2}}
\providecommand{\BIBdecl}{\relax}
\BIBdecl

\bibitem{Cherry1957}
E.~C. Cherry, \emph{{On human communication}}.\hskip 1em plus 0.5em minus 0.4em\relax Cambridge, MA: MIT Press, 1957.

\bibitem{Haykin2005}
S.~Haykin and Z.~Chen, ``{The cocktail party problem},'' \emph{Neural computation}, vol.~17, no.~9, pp. 1875--1902, 2005.

\bibitem{Shinn-Cunningham2008}
B.~Shinn-Cunningham, ``{Object-based auditory and visual attention,},'' \emph{Trends in Cognitive Sciences}, vol.~12, no.~5, pp. 182--186, 5 2008.

\bibitem{Kerlin10a}
\BIBentryALTinterwordspacing
J.~R. Kerlin, A.~J. Shahin, and L.~M. Miller, ``{Attentional Gain Control of Ongoing Cortical Speech Representations in a “Cocktail Party”},'' \emph{The Journal of Neuroscience}, vol.~30, no.~2, pp. 620--628, 1 2010.
\BIBentrySTDinterwordspacing

\bibitem{Mesgarani2012}
\BIBentryALTinterwordspacing
N.~Mesgarani and E.~F. Chang, ``{Selective cortical representation of attended speaker in multi-talker speech perception},'' \emph{Nature}, vol. 485, no. 7397, pp. 233--236, 5 2012.
\BIBentrySTDinterwordspacing

\bibitem{Zatorre1999}
R.~J. Zatorre, T.~A. Mondor, and A.~C. Evans, ``{Auditory attention to space and frequency activates similar cerebral systems},'' \emph{NeuroImage}, vol.~10, no.~5, pp. 544--554, 11 1999.

\bibitem{Petkov2004}
C.~I. Petkov, X.~Kang, K.~Alho, O.~Bertrand, E.~W. Yund, and D.~L. Woods, ``{Attentional modulation of human auditory cortex},'' \emph{Nature neuroscience}, vol.~7, no.~6, pp. 658--663, 6 2004.

\bibitem{Treue1999}
\BIBentryALTinterwordspacing
S.~Treue and J.~C.~M. Trujillo, ``{Feature-based attention influences motion processing gain in macaque visual cortex},'' \emph{Nature}, vol. 399, no. 6736, pp. 575--579, 1999.
\BIBentrySTDinterwordspacing

\bibitem{Saenz2002}
\BIBentryALTinterwordspacing
M.~Saenz, G.~T. Buracas, and G.~M. Boynton, ``{Global effects of feature-based attention in human visual cortex},'' \emph{Nature Neuroscience}, vol.~5, no.~7, pp. 631--632, 7 2002.
\BIBentrySTDinterwordspacing

\bibitem{Liu2011a}
T.~Liu, ``{Constant spread of feature-baed attention across the visual field.}'' \emph{Vision research}, 2011.

\bibitem{OCraven1999}
K.~M. O'Craven, P.~E. Downing, and N.~Kanwisher, ``{fMRI evidence for objects as the units of attentional selection},'' \emph{Nature}, vol. 401, no. 6753, pp. 584--587, 1999.

\bibitem{Schoenfeld2003}
M.~Schoenfeld, ``{Dynamics of feature binding during object-selective attention},'' \emph{Proceedings of the National Academy of Sciences}, 2003.

\bibitem{Wegener2008}
D.~Wegener and F.~Ehn, ``{Feature-based attention and the suppression of non-relevant object features},'' \emph{Vision Research}, vol.~48, no.~27, pp. 2696--2707, 12 2008.

\bibitem{Freeman2014}
E.~Freeman and {Macaluso E}, ``{fmri correlates of object-based attentional facilitation vs. suppression of irrelevant stimuli, dependent on global grouping and endogenous cueing},'' \emph{Frontiers in integrative neuroscience}, 2014.

\bibitem{kravitz2011}
\BIBentryALTinterwordspacing
D.~J. Kravitz and M.~Behrmann, ``{Space-, object-, and feature-based attention interact to organize visual scenes},'' \emph{Attention, Perception, {\&} Psychophysics}, vol.~73, no.~8, pp. 2434--2447, 11 2011.
\BIBentrySTDinterwordspacing

\bibitem{Mayer2012}
\BIBentryALTinterwordspacing
K.~M. Mayer and Q.~C. Vuong, ``{The influence of unattended features on object processing depends on task demand},'' \emph{Vision Research}, vol.~56, pp. 20--27, 3 2012.
\BIBentrySTDinterwordspacing

\bibitem{Wegener2014}
\BIBentryALTinterwordspacing
D.~Wegener, F.~O. Galashan, M.~K. Aurich, and A.~K. Kreiter, ``{Attentional spreading to task-irrelevant object features: experimental support and a 3-step model of attention for object-based selection and feature-based processing modulation},'' \emph{Frontiers in Human Neuroscience}, vol.~8, 6 2014.
\BIBentrySTDinterwordspacing

\bibitem{Geigerman2016}
\BIBentryALTinterwordspacing
S.~Geigerman, P.~Verhaeghen, and J.~Cerella, ``{To bind or not to bind, that's the wrong question: Features and objects coexist in visual short-term memory},'' \emph{Acta Psychologica}, vol. 167, pp. 45--51, 6 2016.
\BIBentrySTDinterwordspacing

\bibitem{Ahveninen2006}
\BIBentryALTinterwordspacing
J.~Ahveninen, I.~P. J{\"{a}}{\"{a}}skel{\"{a}}inen, T.~Raij, G.~Bonmassar, S.~Devore, M.~H{\"{a}}m{\"{a}}l{\"{a}}inen, S.~Lev{\"{a}}nen, F.-H. Lin, M.~Sams, B.~G. Shinn-Cunningham, T.~Witzel, and J.~W. Belliveau, ``{Task-modulated “what” and “where” pathways in human auditory cortex},'' \emph{Proceedings of the National Academy of Sciences}, vol. 103, no.~39, pp. 14\,608--14\,613, 9 2006.
\BIBentrySTDinterwordspacing

\bibitem{Krumbholz2007}
K.~Krumbholz, S.~B. Eickhoff, and G.~R. Fink, ``{Feature- and object-based attentional modulation in the human auditory "where" pathway},'' \emph{Journal of cognitive neuroscience}, vol.~19, no.~10, pp. 1721--1733, 10 2007.

\bibitem{dacosta2013}
\BIBentryALTinterwordspacing
S.~Da~Costa, W.~van~der Zwaag, L.~M. Miller, S.~Clarke, and M.~Saenz, ``{Tuning In to Sound: Frequency-Selective Attentional Filter in Human Primary Auditory Cortex},'' \emph{The Journal of Neuroscience}, vol.~33, no.~5, pp. 1858--1863, 1 2013.
\BIBentrySTDinterwordspacing

\bibitem{Paltoglou2009}
\BIBentryALTinterwordspacing
A.~E. Paltoglou, C.~J. Sumner, and D.~A. Hall, ``{Examining the role of frequency specificity in the enhancement and suppression of human cortical activity by auditory selective attention},'' \emph{Hearing Research}, vol. 257, no. 1-2, pp. 106--118, 2009.
\BIBentrySTDinterwordspacing

\bibitem{Navon1977}
D.~Navon, ``{Forest Before Trees: The Precedence of Global Features in Visual Perception},'' Tech. Rep., 1977.

\bibitem{Oliva2007}
\BIBentryALTinterwordspacing
A.~Oliva and A.~Torralba, ``{The role of context in object recognition},'' \emph{Trends in Cognitive Sciences}, vol.~11, no.~12, pp. 520--527, 12 2007.
\BIBentrySTDinterwordspacing

\bibitem{hedge2008}
J.~HEGDE, ``{Time course of visual perception: Coarse-to-fine processing and beyond},'' \emph{Progress in Neurobiology}, vol.~84, no.~4, pp. 405--439, 4 2008.

\bibitem{Naatanen2007}
\BIBentryALTinterwordspacing
R.~N{\"{a}}{\"{a}}t{\"{a}}nen, P.~Paavilainen, T.~Rinne, and K.~Alho, ``{The mismatch negativity (MMN) in basic research of central auditory processing: A review},'' \emph{Clinical Neurophysiology}, vol. 118, no.~12, pp. 2544--2590, 12 2007.
\BIBentrySTDinterwordspacing

\bibitem{Kaya2014}
\BIBentryALTinterwordspacing
E.~M. Kaya and M.~Elhilali, ``{Investigating bottom-up auditory attention},'' \emph{Frontiers in Human Neuroscience}, vol.~8, p. 327, 5 2014.
\BIBentrySTDinterwordspacing

\bibitem{bohm2013}
T.~M. B{\H{o}}hm, L.~Shestopalova, A.~Bendixen, A.~G. Andreou, J.~Georgiou, G.~Garreau, P.~Pouliquen, A.~Cassidy, S.~L. Denham, and I.~Winkler, ``{The role of perceived source location in auditory stream segregation: Separation affects sound organization, common fate does not},'' \emph{Learning {\&} Perception}, vol.~5, no. Supplement 2, pp. 55--72, 6 2013.

\bibitem{Cherry1953}
E.~C. Cherry, ``{Some experiments on the recognition of speech, with one and with two ears},'' \emph{Journal of the Acoustical Society of America}, vol.~25, no.~5, pp. 975--979, 1953.

\bibitem{Mondor1995}
\BIBentryALTinterwordspacing
T.~A. Mondor and R.~J. Zatorre, ``{Shifting and focusing auditory spatial attention},'' \emph{J Exp Psychol Hum Percept Perform}, vol.~21, no.~2, pp. 387--409, 4 1995.
\BIBentrySTDinterwordspacing

\bibitem{Best2008}
V.~Best, E.~J. Ozmeral, N.~Kop{\v{c}}o, and B.~G. Shinn-Cunningham, ``{Object continuity enhances selective auditory attention},'' \emph{Proceedings of the National Academy of Sciences of the United States of America}, 2008.

\bibitem{Ding2012}
\BIBentryALTinterwordspacing
N.~Ding and J.~Z. Simon, ``{Emergence of neural encoding of auditory objects while listening to competing speakers},'' \emph{Proceedings of the National Academy of Sciences}, vol. 109, no.~29, pp. 11\,854--11\,859, 7 2012.
\BIBentrySTDinterwordspacing

\bibitem{Altmann2008}
\BIBentryALTinterwordspacing
C.~F. Altmann, M.~Henning, M.~K. D{\"{o}}ring, and J.~Kaiser, ``{Effects of feature-selective attention on auditory pattern and location processing},'' \emph{NeuroImage}, vol.~41, no.~1, pp. 69--79, 5 2008.
\BIBentrySTDinterwordspacing

\bibitem{Degerman2008}
\BIBentryALTinterwordspacing
A.~Degerman, T.~Rinne, A.~S{\"{a}}rkk{\"{a}}, J.~Salmi, and K.~Alho, ``{Selective attention to sound location or pitch studied with event‐related brain potentials and magnetic fields},'' \emph{European Journal of Neuroscience}, vol.~27, no.~12, pp. 3329--3341, 6 2008.
\BIBentrySTDinterwordspacing

\bibitem{Fritz2003NatNeur}
\BIBentryALTinterwordspacing
J.~Fritz, S.~Shamma, M.~Elhilali, and D.~Klein, ``{Rapid task-related plasticity of spectrotemporal receptive fields in primary auditory cortex},'' \emph{Nature Neuroscience}, vol.~6, no.~11, pp. 1216--1223, 11 2003.
\BIBentrySTDinterwordspacing

\bibitem{Banerjee2011OscillatoryMechanisms}
S.~Banerjee and A.~Snyder, ``{Oscillatory alpha-band mechanisms and the deployment of spatial attention to anticipated auditory and visual target locations: supramodal or sensory-specific control mechanisms?}'' \emph{Journal of Neuroscience}, 2011.

\bibitem{Frey2014SelectiveTask}
J.~Frey and N.~Mainy, ``{Selective Modulation of Auditory Cortical Alpha Activity in an Audiovisual Spatial Attention Task},'' \emph{Journal of Neuroscience}, 2014.

\bibitem{Bregman1990}
A.~S. Bregman, \emph{{Auditory scene analysis: the perceptual organization of sound}}.\hskip 1em plus 0.5em minus 0.4em\relax Cambridge, Mass.: MIT Press, 1990.

\bibitem{Scholl2001}
B.~Scholl, ``{Objects and attention: the state of the art},'' \emph{Cognition}, vol.~20, no. 1-2, pp. 1--46, 2001.

\bibitem{Busse2005}
L.~Busse and K.~Roberts, ``{The spread of attention across modalities and space in a multisensory object},'' \emph{Proc. Natl. Acad. Sci. U.S.A}, vol. 102, no.~51, pp. 18\,751--18\,756, 2005.

\bibitem{soto2004}
D.~Soto and M.~Blanco, ``{Spatial attention and object-based attention: a comparison within a single task},'' \emph{Vision Research}, vol.~44, no.~1, pp. 69--81, 1 2004.

\bibitem{Maunsell2006}
\BIBentryALTinterwordspacing
J.~H. Maunsell and S.~Treue, ``{Feature-based attention in visual cortex},'' vol.~29, no.~6, pp. 317--322, 2006.
\BIBentrySTDinterwordspacing

\bibitem{Maddox2012}
R.~Maddox and B.~Shinn-Cunningham, ``{Influence of Task-Relevant and Task-Irrelevant Feature Continuity on Selective Auditory Attention},'' \emph{Journal of the Association for Research in Otolaryngology}, vol.~13, pp. 119--129, 2012.

\bibitem{Cusack2004}
\BIBentryALTinterwordspacing
R.~Cusack, J.~Deeks, G.~Aikman, and R.~P. Carlyon, ``{Effects of location, frequency region, and time course of selective attention on auditory scene analysis},'' vol.~30, no.~4, pp. 643--656, 2004.
\BIBentrySTDinterwordspacing

\bibitem{Micheyl2005}
\BIBentryALTinterwordspacing
C.~Micheyl, B.~Tian, R.~P. Carlyon, and J.~P. Rauschecker, ``{Perceptual organization of tone sequences in the auditory cortex of awake macaques},'' \emph{Neuron}, vol.~48, no.~1, pp. 139--148, 2005.
\BIBentrySTDinterwordspacing

\bibitem{Taya2009}
S.~Taya and {Adams WJ}, ``{The fate of task-irrelevant visual motion: perceptual load versus feature-based attention},'' \emph{Journal of vision}, 2009.

\bibitem{Oostenveld2011}
\BIBentryALTinterwordspacing
R.~Oostenveld, P.~Fries, E.~Maris, and J.~M. Schoffelen, ``{FieldTrip: Open source software for advanced analysis of MEG, EEG, and invasive electrophysiological data},'' \emph{Computational Intelligence and Neuroscience}, vol. 2011, p. 156869, 12 2011.
\BIBentrySTDinterwordspacing

\bibitem{Snedecor1989}
G.~Snedecor and W.~Cochran, \emph{{Statistical Methods}}.\hskip 1em plus 0.5em minus 0.4em\relax Iowa State University Press, Ames, 1989.

\bibitem{deCheveigne2018b}
A.~de~Cheveigne and G.~Di~Liberto, ``{Multiway Canonical Correlation Analysis of Brain Signals},'' \emph{NeuroImage}, 2018.

\bibitem{Aiken2008}
\BIBentryALTinterwordspacing
S.~J. Aiken and T.~W. Picton, ``{Human Cortical Responses to the Speech Envelope},'' \emph{Ear and Hearing}, vol.~29, no.~2, pp. 139--157, 4 2008.
\BIBentrySTDinterwordspacing

\end{thebibliography}

\end{document}